\documentclass[11pt]{article}
\usepackage{moriond}
\usepackage{subfigure} 
\usepackage[pdftex]{graphicx} 
\pdfoutput=1

\begin{document}
\vspace*{4cm}
\title{HEAVY FLAVOUR IN A NUTSHELL}

\author{ Robert W. Lambert }

\address{CERN, Geneva, Switzerland}

\maketitle\abstracts{Moriond QCD brings together particle physicists of varied interests. This review and introduction to heavy flavour physics is aimed at those not in the heavy-flavour field to describe the motivation and methodology of precision flavour physics, and introduce the most tantalising searches for new physics. The LHC experiments are expected to make great inroads into constraining the new physics parameter space and discover the new physics which I will argue \emph{must} be present to describe our observed universe.}

\section{Introduction}
\label{Section:intro}

Heavy flavour is a broad subject both experimentally and theoretically, stretching back two hundred years to the proposal of the first flavoured object, the proton, in 1815~\cite{Prout:Proton}. In this paper the general topics and basic theory aspects are discussed as needed to develop an understanding of the field today, such that the Reader may be equipped to understand the remaining proceedings from this section of the conference and participate in discussions with their colleagues over the key results.

This paper is a summary of existing works, particularly three very interesting and important papers of the last twelve months: the measurement by the D{\O} collaboration of a $3.2\sigma$ deviation from the Standard Model in the flavour-specific asymmetry of neutral $B$-meson mixing~\cite{D0:mumu:2010}, an update of $B$-mixing both theoretically and experimentally by Lenz and Nierste~\cite{Nierste:Bmix:2011}, plus the recently updated results of the WMAP seven-year sky survey~\cite{WMAP7}. 

For more complete, more advanced, theoretical and experimental summaries, the Reader is directed to the similar introduction in the previous conference in this series~\cite{Wilkinson:Moriond:2010}, the summary paper mentioned earlier~\cite{Nierste:Bmix:2011}, and the other excellent proceedings from this same conference.

\subsection{Welcome to Our Universe}
\label{Section:intro:Universe}

The Wilkinson Microwave Anisotropy Probe, WMAP, makes precision measurements of the properties of the cosmic microwave background radiation. It is amazing that such a simple experiment can produce some of the most profound results in physics. In combination with some other simple cosmological observations, 
 we are able to measure the cosmological density of the universe and divide the total density into constituents from different sources. Shockingly we discover only 5\,\% of the mass of the universe can be attributed to baryonic matter~\cite{WMAP7}. 
 However, even 5\,\% is much higher than could be expected from our existing theories which stipulate all matter and antimatter should have been annihilated just after the big bang. 

The fact there is any matter left over at all requires there to be a difference between the behaviour of matter and antimatter, which violates CP-symmetry (CPV)~\cite{Sakharov}. We can quantify the amount of CPV required by taking the ratio of the remaining matter to the number of photons from the annihilation, and we find this number is approximately one part in a billion. In our Standard Model of particle physics (SM), even with maximal CPV, the equivalent predicted asymmetry is lower than the observed asymmetry by more than ten orders of magnitude. We know then that there must be new physics waiting to be discovered and that it must contain new sources of CPV.

There are indeed many other problems with what we call the Standard Model; but the puzzles of dark matter and dark energy, the matter-antimatter asymmetry, and the lack of a good quantum theory of gravity, are the most obvious shortfalls.

\subsection{Heavy Flavour as a Tool for New Physics}
\label{Section:intro:HF}

Heavy flavour physics is a precision tool to discover new physics. The reach of heavy flavour is very broad since the production and decay of any heavy meson inevitably involve aspects from every portion of the Standard Model. Even a simple decay such as $B_s^0\to{D_s^{(*)(*)\pm}l^{\mp}\nu_l}$, probes: all three generations of quarks and leptons, QCD, QED, and weak interactions. Arguably also the Higgs mechanism and even the top quark (mediating the observed $B_s^0$-mixing) play a role. Heavy flavour is a microcosm of the entire Standard Model so it should exhibit the same flaws as the Standard Model and probe all avenues of new physics.

Precision measurements are completely complementary to direct searches for new physics. Direct searches for new heavy particles at the energy frontier are limited in their reach by the energy of the collider at hand. Precision measurements, however, are sensitive to the quantum-mechanical effects of new physics in loops and virtual processes, to scales well beyond the energy of the collider. Typically we say up to 1000-times the energy of the collider.


Having identified that heavy flavour is a powerful tool to search for new physics, we would like to use it to answer the following two questions.
\begin{enumerate}
  \item Where is the CP-violation we need to explain the observed universe?
  \item Given that there must be new physics, what is its flavour structure?
\end{enumerate}

To answer those questions we must identify how and where to look for new physics, and for that we need a recipe or a map.

\subsection{A Map of the Search for New Physics}
\label{Section:intro:Map}

When looking for new physics we can follow the following prescription. Identify channels and observables where new physics is not expected and make precision measurements of well-predicted observables. Then identify a similar or related area where new physics can enter and perform precision measurements of related observables to detect any new physics.

Tree-level Standard-Model-like decays are a good example of where new physics is not expected. Consequently to look for new physics we are especially interested in channels with loops and penguins (radiative loops), where any new physics charges, currents, and virtual heavy particles, can enter into the loop and change the result dramatically.

We are also interested in looking for new sources of CP-violation. In the SM there is only one source of CP-violation, which is a phase in the weak mixing matrix, the CKM matrix. To observe this phase, or any new physics phase, we construct observables with two competing phases, and measure phase differences through interference. In the Standard Model the CKM phase manifests most obviously in the $b$-quark system~\cite{BCPV}, which again emphasises that heavy flavour physics is crucial. We can construct many different observables to measure this single known phase, and search closely for inconsistencies, the signs of new CP-violating physics.

\subsection{Status of the CKM-mechanism}
\label{Section:Status}

\begin{figure}[btp]
\centering
\subfigure{\includegraphics[width=0.35\textwidth,keepaspectratio,]{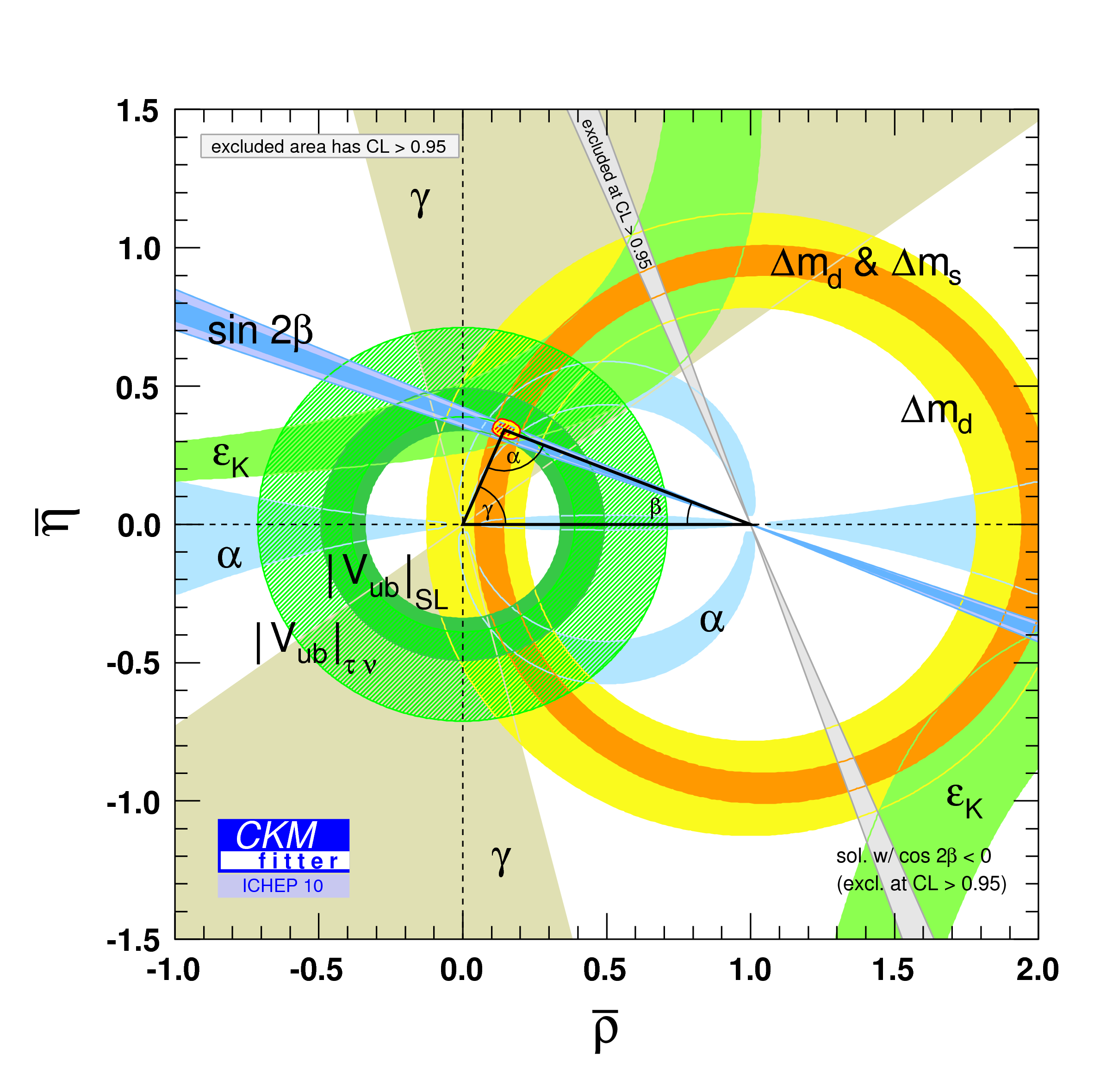}}\hspace{0.02\textwidth}
\subfigure{\includegraphics[width=0.55\textwidth,keepaspectratio,]{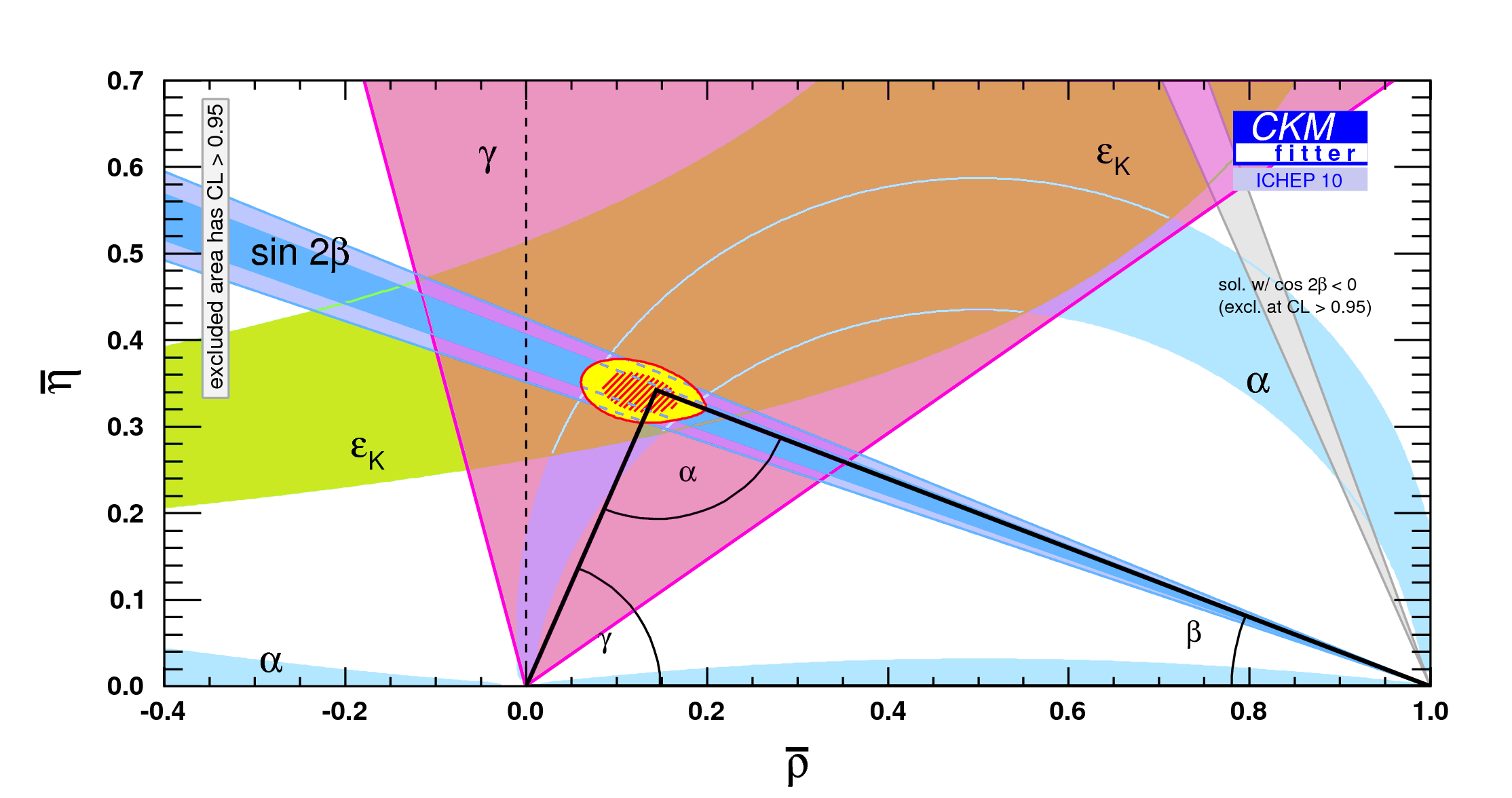}}
\caption{Overlapping constraints on the phase in the CKM matrix, reproduced~\protect\cite{CKM:UT:2007}. $\bar{\rho}$ and $\bar{\eta}$ are effectively the real and imaginary parts of the CKM phase. The red hashed region of the global combination corresponds to 68\,\% CL. On the left all current experimental constraints are used. On the right only observables which explicitly violate CP are used. The CKM-angle $\gamma$ is highlighted in pink to demonstrate it is a weak constraint.}
\label{Figure:UT}
\end{figure}

Combining measurements of CKM-parameters from many different sources we then usually plot all the phase constraints on a single 2D complex plane to constrain the real part ($\bar{\rho}$) and imaginary part ($\bar{\eta}$) of the phase in the CKM~\cite{CKM:UT:2007}. This popular image is reproduced here as Fig.~\ref{Figure:UT}. In the wide range of experimental observables across many different channels, all of the results agree very well and are very consistent with the CKM-model for weak mixing and CP-violation. This confirms that the Standard Model is very self-consistent, and that the CKM-mechanism is an excellent description, but it does not leave much breathing room for new physics.

Fortunately we do have several unexplored and promising places to search for new physics~\cite{LHCb:Roadmap:2009}.

\section{Hottest New Physics Searches}
\label{Section:HNP}

There are many searches for new physics in precision flavour physics, but this paper only briefly covers five areas which are typical of certain classes of precision search.
\begin{enumerate}
  \item Precision CKM-measurements, such as the determination of the CKM-angle $\gamma$.
  \item Decays with penguins and loops, such as the rare decay $B_d^0{\to}K^*\mu^+\mu^-$.
  \item Very rare decays with possible new physics enhancements, such as the rare decays $B_{s/d}^0{\to}\mu^+\mu^-$.
  \item Generic CP-asymmetry searches, such as $B{\to}K\pi$, where we have a so-called ``$K\pi$-puzzle.''
  \item Mixing of heavy neutral mesons, for example $B^0_{s/d}$-mixing.
\end{enumerate}

\subsection{Determination of the CKM-angle {$\gamma$}}

In Fig.~\ref{Figure:UT}, right, is replotted the constraints on the CPV phase, with only the explicit CP-violating observables. Any disagreement in such a plot could point immediately to new CP-violating physics. Here the CKM-angle $\gamma$ is not well known and could hide moderate new physics. More precise measurements of $\gamma$ are planned at the LHC, specifically at LHCb~\cite{LHCb:Roadmap:2009}.

\subsection{{$B_d^0{\to}K^*\mu^+\mu^-$}}

The flavour structure of new physics may be exposed in departures from the SM in penguin and loop processes. $B_d^0{\to}K^*\mu^-\mu^+$ is a rare-decay channel with both penguin \emph{and} other competing loop contributions~\cite{LHCb:Roadmap:2009}. In this channel there are several observables with high sensitivity to new physics, particularly to the angular structure of new physics models such as supersymmetry. One key observable is the forward-backward-asymmetry, $A_{fb}$ where there are many current results available, but where none as yet show a deviation from the Standard Model.

\subsection{{$B_{s/d}^0{\to}\mu^+\mu^-$}}

Extremely rare decays are often very sensitive to new physics contributions. $B_{s/d}^0{\to}\mu^+\mu^-$ are two channels with a very clear experimental signature, very precise theoretical prediction, and very large sensitivity to new physics~\cite{LHCb:Roadmap:2009}. In certain supersymmetric models the SM branching fraction may be increased by a very large factor. Very recent results from the Tevatron and LHCb show no departure so far, but the prospect is good for the coming months~\cite{LHCb:bsmm:2011}.

\subsection{The {$K\pi$}-puzzle}

\begin{figure}[btp]
\centering
\includegraphics[width=0.55\textwidth,keepaspectratio,]{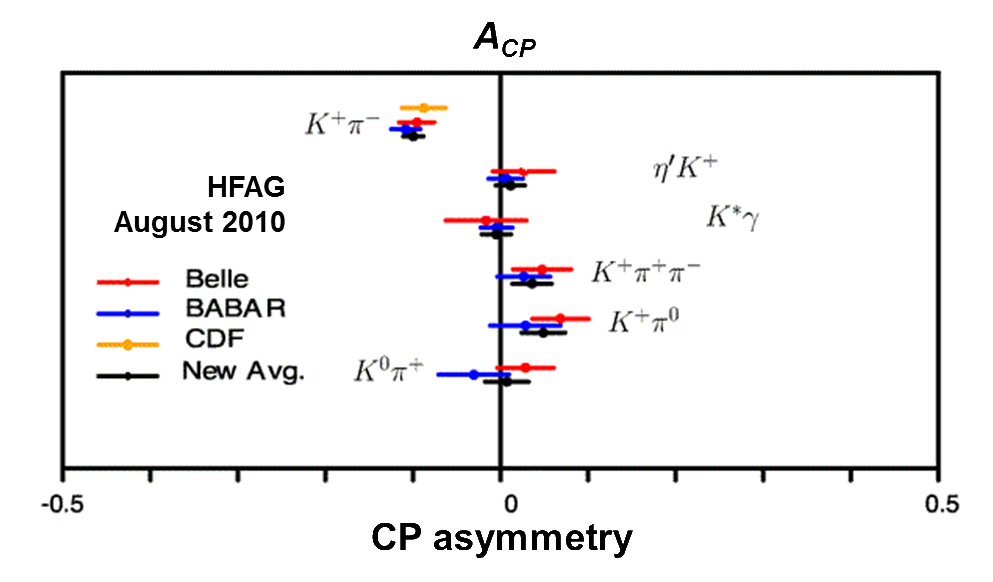}
\caption{Selected measurements of direct CP-asymmetry in decays of the form $B\to{K}\pi$, reproduced~\protect\cite{HFAG:Kpi:2010} with modification. The CP-asymmetry in the $B^0\to{K^+}{\pi^-}$ mode disagrees with all other measurements, this unexpected feature is known as the ``$K\pi$-puzzle.''}
\label{Figure:Kpi}
\end{figure}

Generic searches for observable CP-violation in decays could reveal unexpected new sources of CP-violation. In channels of the form $B\to{K}\pi$ we already have a hint of departure from the expectations. Naively we would expect all decays of this form to exhibit similar levels of direct CP-violation, however, as shown in Fig.~\ref{Figure:Kpi} the CP-asymmetry in the $B^0\to{K^+}{\pi^-}$ mode disagrees with all other measurements. This is known as the ``$K\pi$-puzzle,'' and is an interesting hint for new physics~\cite{HFAG:Kpi:2010}. Precision studies at the LHC will confirm or deny this disagreement~\cite{LHCb:Roadmap:2009}.

\subsection{New Physics in Mixing}

\begin{figure}[btp]
\centering
\subfigure{\includegraphics[width=0.40\textwidth,keepaspectratio,]{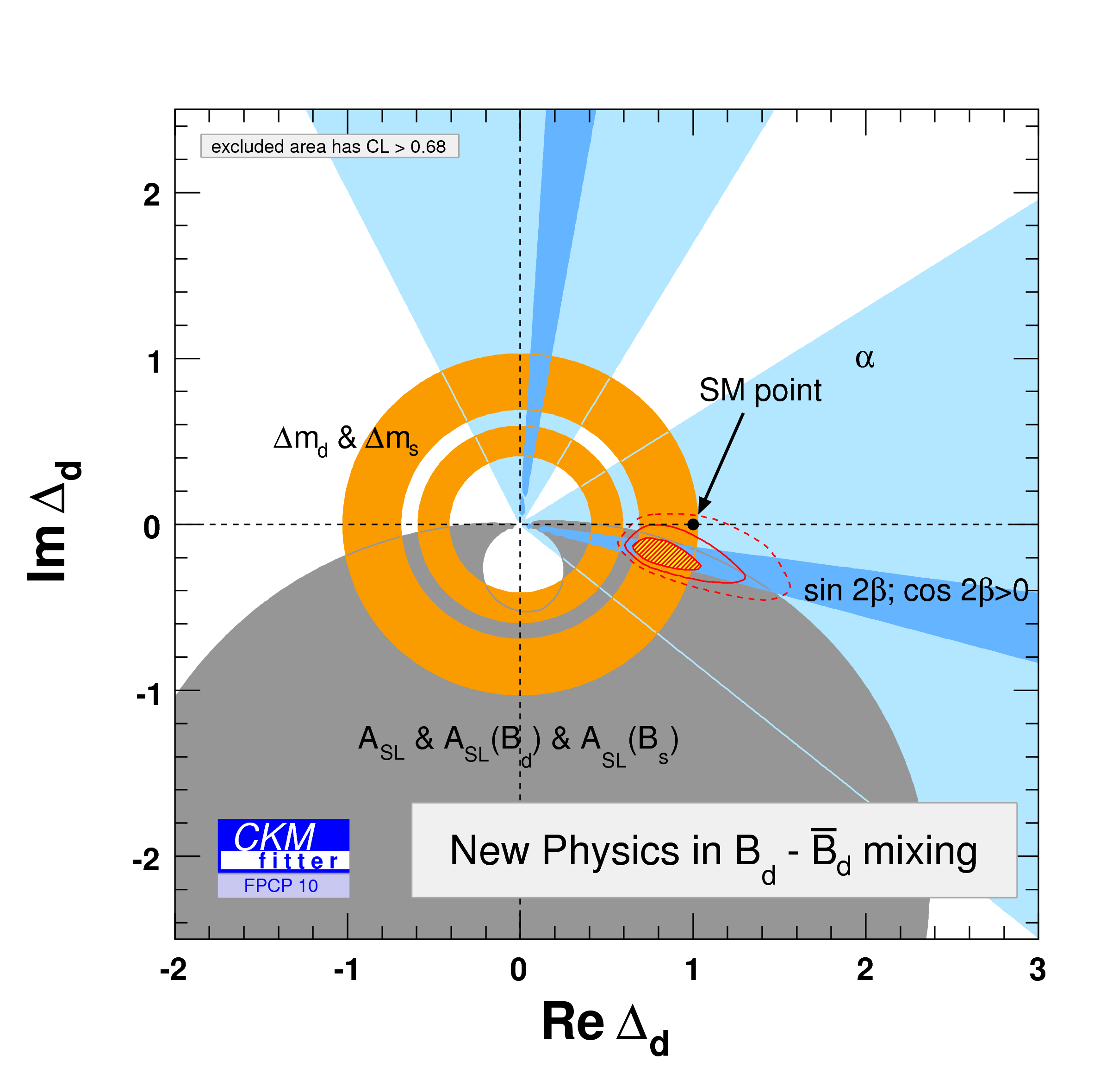}}\hspace{0.05\textwidth}
\subfigure{\includegraphics[width=0.40\textwidth,keepaspectratio,]{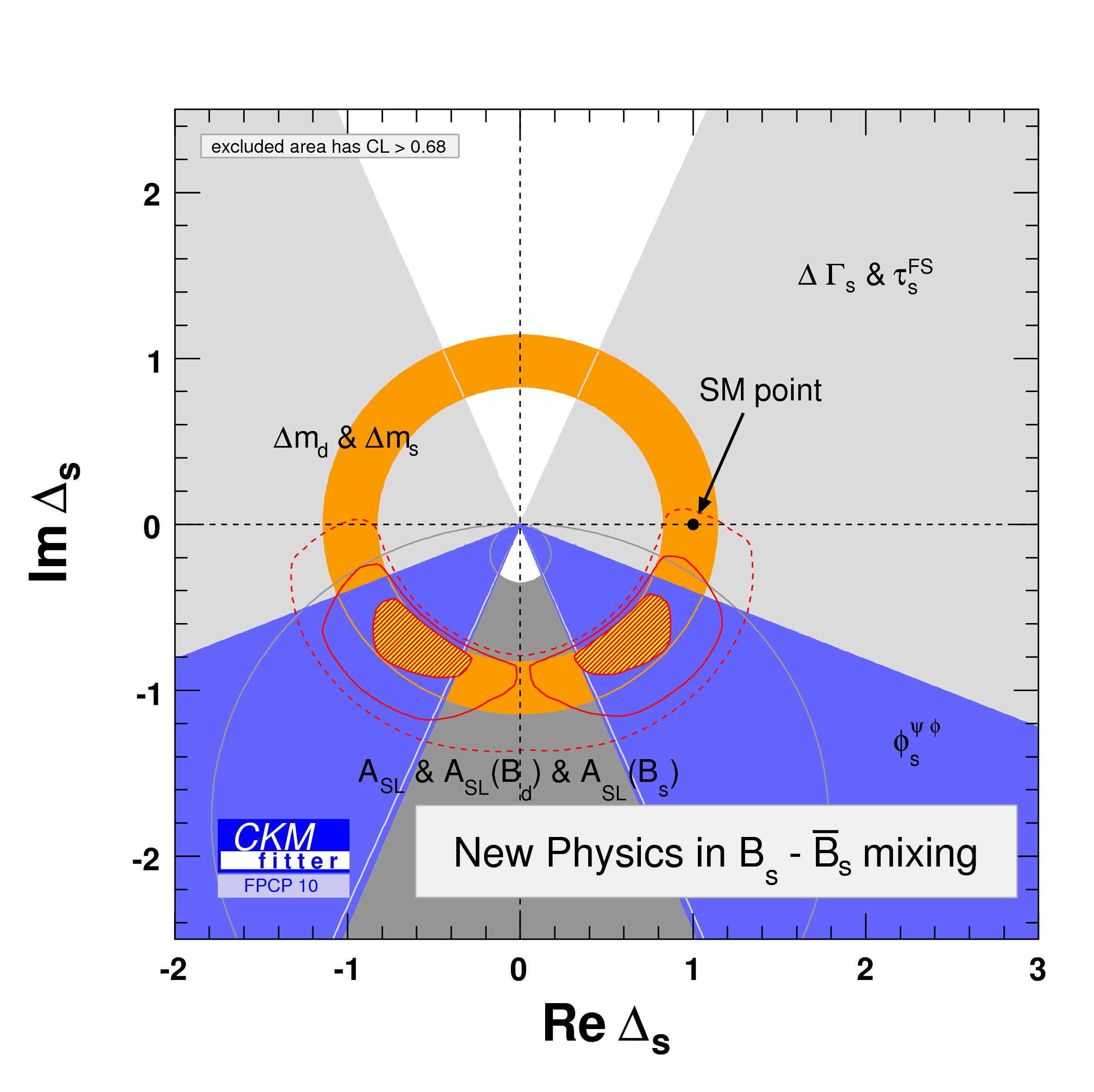}}
\caption{Constraints on new physics in neutral $B$-meson mixing, for the matrix element $M_{12}$, reproduced~\protect\cite{Lenz:BmixFull:2011}. $\Delta_q=(M^{NP}_{12}/M^{SM}_{12})$, $q=s,d$. A combined fit finds that the Standard Model point at (1,0) is disfavoured by 3.6$\sigma$, which rests mostly on the recent measurement of flavour-specific asymmetry by the D{\O} collaboration~\protect\cite{D0:mumu:2010}.}
\label{Figure:NPMixing}
\end{figure}

Recently it has been reported that physics in neutral $B$-meson mixing is already divergent from the Standard Model~\cite{Lenz:BmixFull:2011} by more than $3\sigma$. Mixing is a very curious, unintuitive, pure quantum-mechanical phenomenon, where particle and antiparticle partners both contribute to the same observed state. The observed state is an oscillating time-dependent mixture of particle and antiparticle, and the oscillation is mediated by a box diagram in the SM, most simply described by a mixing matrix. Since mixing is a loop-level process, generic new physics can change both the magnitude and the phase of the mixing, and so it is usual to define a complex number parameter to characterise the new physics contribution. In the selected analysis~\cite{Lenz:BmixFull:2011} the authors choose to allow for new physics only in the most sensitive element of the mixing matrix, $M_{12}$, and so define $\Delta_q=(M^{NP}_{12}/M^{SM}_{12})$, the complex ratio of the new physics and Standard Model values. This parameter is constrained by several current measurements as reproduced here in Fig.~\ref{Figure:NPMixing}. The measurements currently agree, but with a central value which is $3.6\sigma$ from the SM. 

The majority of this departure can be attributed to the recent measurement by the D{\O} collaboration~\cite{D0:mumu:2010}, the first independent evidence for new CPV physics.

\section{Flavour-Specific Asymmetry}
\label{Section:Afs}

The D{\O} collaboration recently produced an exciting and surprizing result in the measurement of flavour-specific asymmetry in the semileptonic decays of $b$-quarks \cite{D0:mumu:2010}. They determine the total dimuon charge asymmetry, which is interpreted as the direct result of the flavour-specific asymmetries in the $B_s^0$ and $B_d^0$ system ($a_{fs}^s$ and $a_{fs}^d$, respectively). They measure a quantity $$A^{b}{\approx}(a_{fs}^{s}+a_{fs}^{d})/2=[-9.57{\pm}2.51(stat){\pm}1.46(syst)]{\times}10^{-3}$$ which is 3.2 standard deviations from the Standard Model prediction \cite{D0:mumu:2010}.

\begin{figure}[btp]
\centering
\subfigure{\includegraphics[width=0.40\textwidth,keepaspectratio,]{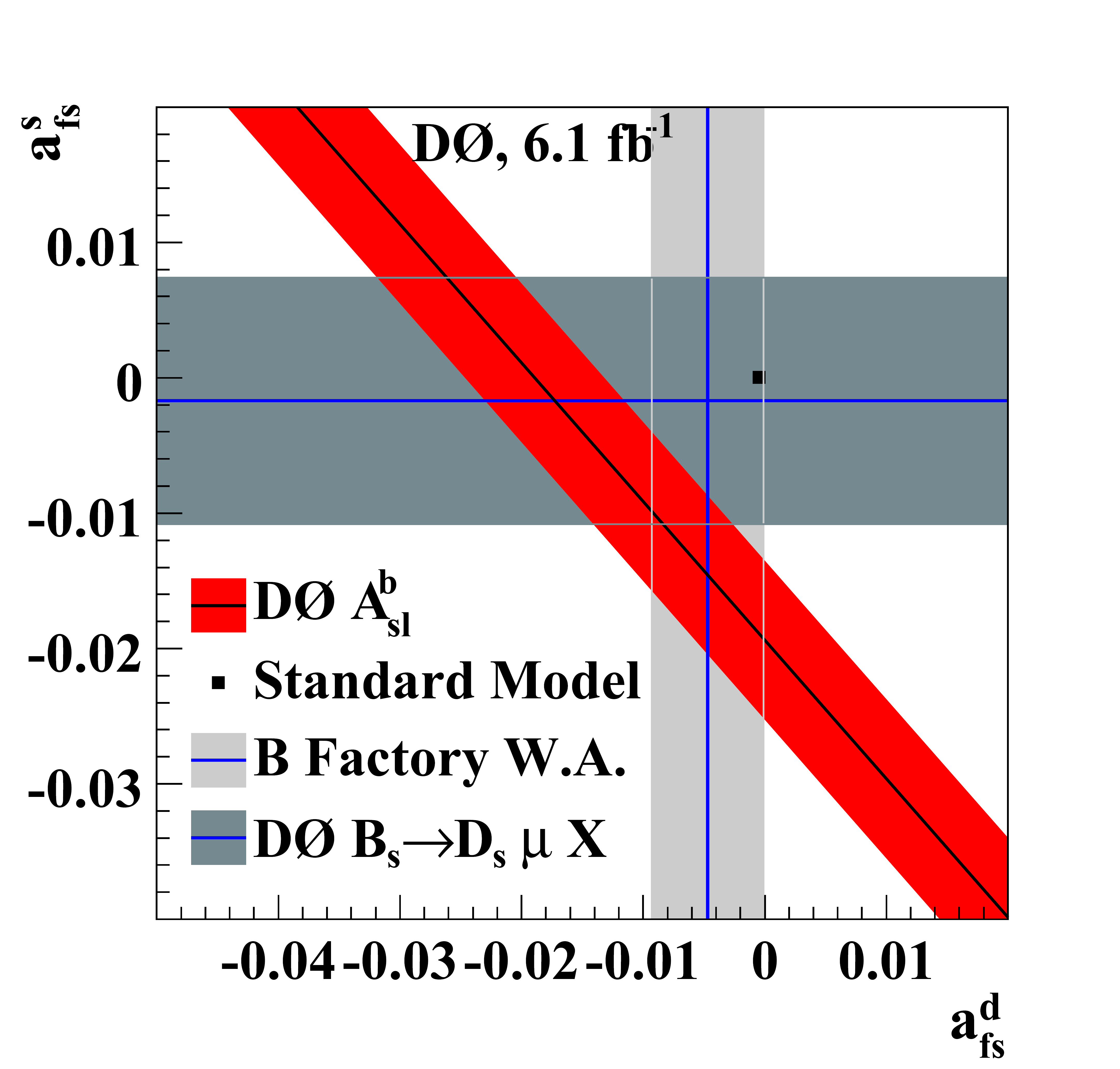}}\hspace{0.05\textwidth}
\subfigure{\includegraphics[width=0.40\textwidth,keepaspectratio,]{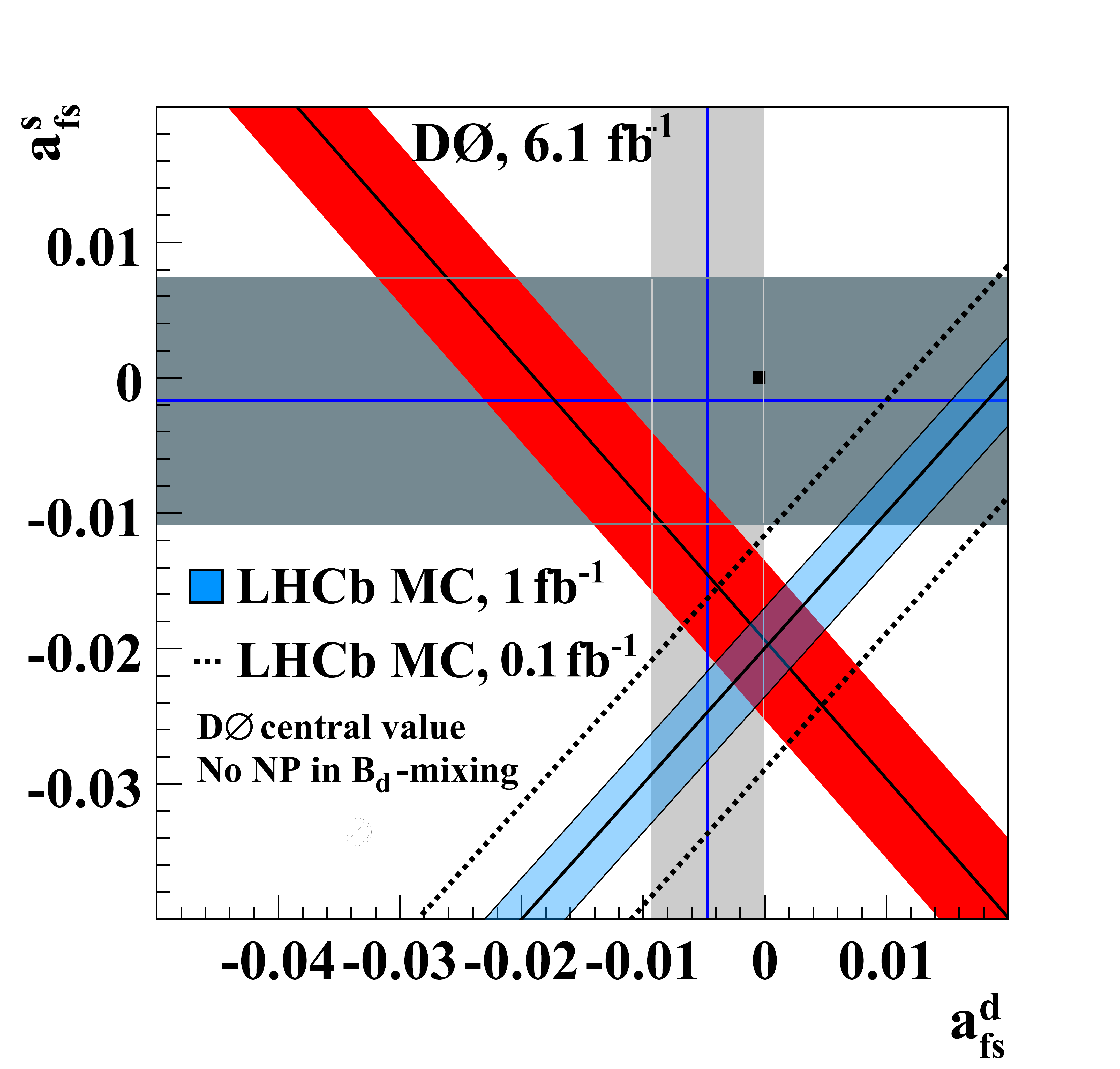}}
\caption{Measurements and prospects for new physics in flavour-specific asymmetry. The asymmetry in neutral $B^0$ mixing, $a_{fs}^d$ is plotted against the asymmetry in neutral $B^0_s$-mixing, $a_{fs}^s$. The left figure is reproduced~\protect\cite{D0:mumu:2010} (slightly modified), the recent D{\O} measurement in red is the first measurement inconsistent with the Standard Model point of $\sim(0,0)$. The right figure also has superimposed the LHCb expected result from simulation (Monte Carlo), should the D{\O} central value hold and should there be no new physics in $a_{fs}^d$. There we scale the Monte Carlo (MC) prediction~\protect\cite{RWL:Thesis} to the yields in real data and add also the expected systematic uncertainties. }
\label{Figure:Afs}
\end{figure}

Fig.~\ref{Figure:Afs} is reproduced~\cite{D0:mumu:2010} and slightly modified to also include the expected LHCb sensitivity taken from simulation (Monte Carlo or MC), applying the real-data yields and estimates of systematic uncertainties. In the environment of the LHC, such a measurement is made more challenging by the expected production asymmetry \cite{Nierste:afs:2006}, however, using a novel time-dependent~\cite{RWL:Thesis} technique LHCb can make an accurate measurement of ${\Delta}A_{fs}=(a_{fs}^{s}-a_{fs}^{d})/2$, with a statistical sensitivity (as predicted from the MC) of $\approx2{\times}10^{-3}$ in 1\,fb$^{-1}$. This measurement is complementary to the D{\O} measurement, and almost orthogonal in the $(a_{fs}^s:a_{fs}^d)$-plane. 

\section{Conclusion}
\label{Section:Conclusion}

I have argued that there must be new physics waiting to be discovered such that our particle physics theory can describe the observed universe. The LHC is a machine purpose-built to discover this new physics. At the LHC the complementarity of direct searches and precision measurements is crucial to identify and classify the new physics. LHCb is \emph{the} precision heavy-flavour experiment at the LHC and will measure many different observables which all place good constraints on this new physics. Finally we have already seen evidence for a departure of observation from the Standard Model in the mixing of neutral mesons, thanks to the recent measurement from D{\O}. In this interesting area LHCb will endeavour to make an early complementary measurement. We stand at the very beginning of the LHC era, which is already proving to be one of the most exciting times in the history of particle physics.

\section{Acknowledgements}
\label{Section:Ack}

Many thanks to the conference organisers for the invitation. Thanks to J.~Albrecht, U.~Kerzel, T.~Ruf and G.~Wilkinson, for their invaluable support. Thanks also to the CKM fitter group for updating the fit results in the so-called ``$B_s$-triangle,'' pointing out to me a long-standing goof in the LHCb TDR and other publications, including my own Thesis, also for putting up with my crazy questions about their fitting methods.


\section*{References}
\begin{thebibliography}{10}

\bibitem{Prout:Proton}
W.~Prout, 
  Annals of Philosophy
  \textbf{6} (1815) pp.~321-330.

\bibitem{D0:mumu:2010}
D{\O} Collaboration, 
  Phys.~Rev.~Lett.~{\bf 105} (2010)
  pp.~081801, hep-ex arxiv:1005.2757.

\bibitem{Nierste:Bmix:2011}
A.~Lenz and U.~Nierste, 
  hep-ph arXiv:1102.4274 (2010).

\bibitem{WMAP7}
N.~Jarosik \emph{et al.}, 
  ApJS~ \textbf{192} 14 (2011).

\bibitem{Wilkinson:Moriond:2010}
G.~Wilkinson, proceedings of this conference, 2010.

\bibitem{Sakharov}
A.~D.~Sakharov, 
  JETP \textbf{5} (1967) pp.~24-27.~Republished in Soviet Physics
  Uspekhi.

\bibitem{BCPV}
A.~B.~Carter and A.~I.~Sanda, 
  Phys.~Rev.~D~\textbf{23} (1981) pp.~1567-1579.

\bibitem{CKM:UT:2007}
CKMfitter Group, 
  Eur.~Phys.~J.~C41, 1-131 (2005), http://ckmfitter.in2p3.fr.

\bibitem{LHCb:Roadmap:2009}
LHCb collaboration, 
  CERN-LHCb-PUB-2009-029, hep-ex arXiv:0912.4179.

\bibitem{HFAG:Kpi:2010}
HFAG, http://www.slac.stanford.edu/xorg/hfag/rare/ichep10/acp/index.html

\bibitem{LHCb:bsmm:2011}
LHCb collaboration, 
  CERN-LHCb-PH-EP-2011-029, hep-ex arXiv:1103.2465.

%
%
%

\bibitem{Lenz:BmixFull:2011}
A.~Lenz \emph{et al.}, 
  Phys.~Rev.~D~\textbf{83} (2011) pp.~036004, hep-ph arXiv:1102.4274.

\bibitem{Nierste:afs:2006}
U.~Nierste, 
  hep-ph 0406300 (2006).

\bibitem{RWL:Thesis}
R.~W.~Lambert, 
  CERN-THESIS-2009-001 (2008).

\end{thebibliography}


\end{document}